\documentclass{article}\usepackage{knitr}

\usepackage[margin=.75in]{geometry}
\usepackage{graphicx, ctable, booktabs}
\usepackage{color}
\usepackage{listings}
\usepackage{ltablex, calc, enumerate, multirow, float, soul, paralist}
\usepackage{hyperref}
\usepackage{longtable, pdflscape, textcase}
\usepackage[backend=bibtex, natbib=true, style=authoryear]{biblatex}
\IfFileExists{upquote.sty}{\usepackage{upquote}}{}
\begin{document}

\setlength{\parskip}{3ex}
\setlength{\parindent}{0pt}

\title{Putting Down Roots: A Graphical Exploration of Community Attachment}
\author{Andee Kaplan \and Eric Hare}

%
\date{}

\maketitle

\begin{abstract}
In this paper, we explore the relationships that individuals have with their communities. This work was prepared as part of the ASA Data Expo `13 sponsored by the Graphics Section and the Computing Section, using data provided by the Knight Foundation Soul of the Community survey. The Knight Foundation in cooperation with Gallup surveyed 43,000 people over three years in 26 communities across the United States with the intention of understanding the association between community attributes and the degree of attachment people feel towards their community. These include the different facets of both urban and rural communities, the impact of quality education, and the trend in the perceived economic conditions of a community over time. The goal of our work is to facilitate understanding of why people feel attachment to their communities through the use of an interactive and web-based visualization. We will explain the development and use of web-based interactive graphics, including an overview of the R package \texttt{Shiny} and the JavaScript library \texttt{D3}, focusing on the choices made in producing the visualizations and technical aspects of how they were created. Then we describe the stories about community attachment that unfolded from our analysis.

\end{abstract}

\clearpage

\setcounter{page}{1}
\section{Introduction}

This work was part of the American Statistical Association's Data Exposition 2013. The dataset came from the Knight Foundation's `Soul of the Community' project \citep{SOTC}. For this project, the Knight Foundation in conjunction with Gallup collected surveys from 43,000 people in the years 2008, 2009, and 2010, in 26 U.S. communities. A map of the 26 communities and the geographical regions to which we assigned them is displayed in Figure \ref{fig:region_map}. The regions are the Great Plains, the West, the Deep South, the Southeast, and the Rust Belt. These regions were not included as part of the dataset, but rather were our own constructs created by a graphical exploration of the locations of each community. We roughly based these regions on the US Census regions and divisions of the United States map \citep{Census}. We did not strictly adhere to state boundaries, but rather looked at the proximity of individual cities to the surrounding communities.

\begin{knitrout}
\definecolor{shadecolor}{rgb}{0.969, 0.969, 0.969}\color{fgcolor}\begin{figure}[H]

{\centering \includegraphics[width=\textwidth]{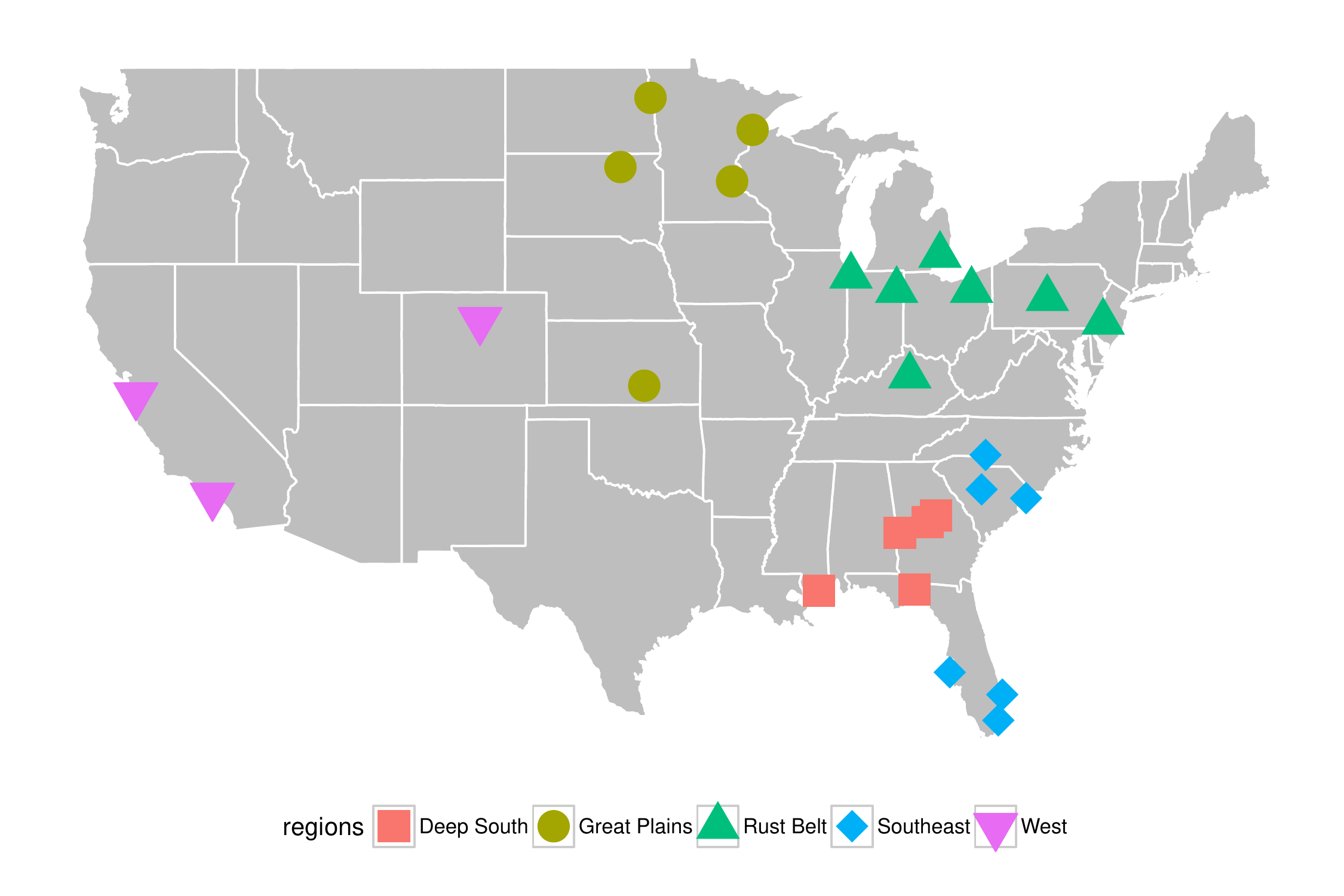} 

}

\caption[Overview of the 26 Knight Foundation communities in which surveys were conducted]{Overview of the 26 Knight Foundation communities in which surveys were conducted. Assigned geographical regions are indicated by both color and shape. The Knight Foundation in conjunction with Gallup collected surveys from 43,000 people in the years 2008, 2009, and 2010 with the goal of understanding the association between community attributes and the degree of attachment people feel towards their community.}\label{fig:region_map}
\end{figure}

\end{knitrout}

The survey contained raw responses as well as derived metrics that we used to gain insight into what makes a community thrive. The metrics we used can be found in Table~\ref{tab:metrics}. Each metric was calculated as a simple average of the response to anywhere from 2 to 6 questions. The metrics gave insight into how residents felt their community rated on various dimensions. For example, Education covered both public education as well as higher education in the community, while Social Offerings dealt with both nightlife as well as neighborliness. Community Attachment combined questions on how proud residents were to live in their community, if they would have recommended the community as a place to live, and how they predicted the community would be in five years. A score of 5 indicated the most positive response was given on all questions that this metric is derived from. We used the 10 metrics in Table~\ref{tab:metrics} to find relationships within types of communities to Community Attachment, as well as to explore any notable regional differences.

\begin{table} \small
\centering
\begin{tabular}{l p{.65\textwidth}}
\hline
Metrics & \\
\hline
\multirow{5}{*}{Community Attachment} &  I am proud to say I live in [Community]. \\
& [Community] is the perfect place for people like me. \\
& Taking everything into account, how satisfied are you with [Community] as a place to live? \\
& How likely are you to recommend [Community] to a friend or associate as a place to live? \\
& And thinking about five years from now, how do you think [Community] will be as a place to live compared to today? \\
\hline
\multirow{3}{*}{Social Offerings} & Having a vibrant nightlife with restaurants, clubs, bars, etc. \\
& Being a good place to meet people and make friends \\
& How much people in [Community] care about each other \\
\hline
\multirow{5}{*}{Openness} & Young, talented college graduates looking to enter the job market \\ 
& Immigrants from other countries \\
& Families with young children \\
& Gay and lesbian people \\
& Senior citizens \\
\hline
\multirow{2}{*}{Aesthetics} & The availability of outdoor parks, playgrounds, and trails \\
& The beauty or physical setting \\
\hline
\multirow{2}{*}{Education} & The overall quality of public schools in your community \\
& The overall quality of the colleges and universities \\
\hline
\multirow{3}{*}{Basic Services} & The highway and freeway system \\
& The availability of affordable housing \\
& The availability and accessibility of quality healthcare\\
\hline
\multirow{2}{*}{Leadership} & The leadership of the elected officials in your city \\
& The leaders in my community represent my interests \\
\hline
\multirow{6}{*}{Economy} & The availability of job opportunities \\
& How would you rate economic conditions in [Community] today? \\
& Right now, do you think that economic conditions in [Community] as a whole are getting better or getting worse? \\
& How likely are you to agree that your job provides you with the income needed to support your 
family? \\
& Now is a good time to find a job in my area \\
& How satisfied are you with your job, that is, the work you do? \\
\hline
\multirow{2}{*}{Safety} & How would you rate how safe you feel walking alone at night within a mile of your home? \\
& How would you rate the level of crime in your community? \\
\hline
\multirow{4}{*}{Social Capital} & How many formal or informal groups or clubs do you belong to, in your area, that meet at least monthly? \\
& How many of your close friends live in your community? \\
& How much of your family lives in this area? \\
& How often do you talk to or visit with your immediate neighbors? \\
\hline
\multirow{4}{*}{Civic Involvement} & Performed local volunteer work for any organization or group \\
& Attended a local public meeting in which local issues were discussed \\
& Voted in the local election \\
& Worked with other residents to make change in the local community\\
\hline 
\end{tabular}
\caption{\label{tab:metrics} The metrics used from the Knight Foundation Soul of the Community survey \citep{SOTCSC}. All metrics are on a 1-3 scale except for Community Attachment, which is on a scale of 1-5. A higher score on any metric indicates the respondent replied positively to the associated questions.}
\end{table}

The goal of our work was to facilitate understanding of why people feel attachment to their communities through the use of an interactive and web-based visualization. Specifically, we took the point of view of a community planner, either from one of the communities in the study or from a community in the same region or a similar urbanicity. By putting the user in the driver seat of their own experience, we allow the user to apply the conclusions of their interaction to their own situation. The purpose of interaction is to discover what the data has to tell the world. We did not attempt to draw statistical conclusions about the data, thus we did not use the survey weights provided in the data in our analysis. Because the communities are sparsely and unevenly distributed throughout the United States, we felt an exploratory approach would help us to sift through the data and discover its patterns.

Many of the discoveries in the data were readily apparent, while others required some more investigation. In the words of John Tukey, ``Exploratory data analysis is detective work - numerical detective work - or counting detective work - or \emph{graphical detective work}." \citep{tukey77} Dynamic, interactive visualizations can empower people to explore the data for themselves as well as encourage engagement with the data in a way that static visualizations cannot. Additionally, linking multiple visualizations shows different aspects of a complex data set and helps highlight relationships. By allowing actions in one plot to affect elements in other plots, comparisons are made easy for the user without requiring much memorization. This aids in pattern finding by reducing cognitive load. In addition to wanting it to be easy to explore the data, we wanted the tool to be easy to use. A web-based application is platform-independent and allows the user to employ the tool without any software to download. Additionally, by building an application that works on all modern browsers and operating systems, there are no limitations on who can use the tool. Finally, automatic feature additions and bug fixes can be completed transparently to the user.

To fully engage the user with our work and facilitate the emergence of interesting or descriptive patterns we created CommuniD3 (available at \url{http://andeek.shinyapps.io/CommuniD3}), an interactive web-based tool that relies heavily on the idea of linked plots. A linked plot will adapt to changes made in other plots within the collection, creating a dynamic and interactive set of graphics. Different visualizations illustrate different aspects of the data, and linking helps regain the multidimensional aspect of the data \citep{buja1991interactive}. In the following section we discuss the structure and tools used to build CommuniD3. Section 3 highlights an application of CommuniD3 in finding interesting stories across the Unitied States.

\section{A Graphical User Interface}

To visually understand attachment, we created an interactive web-based application. The construction and design of CommuniD3 are detailed in the following sections. First the user interface is described and then we discuss the wide range of technology we used to construct the tool.

\begin{figure}[H]
\centering
\includegraphics[width=\textwidth]{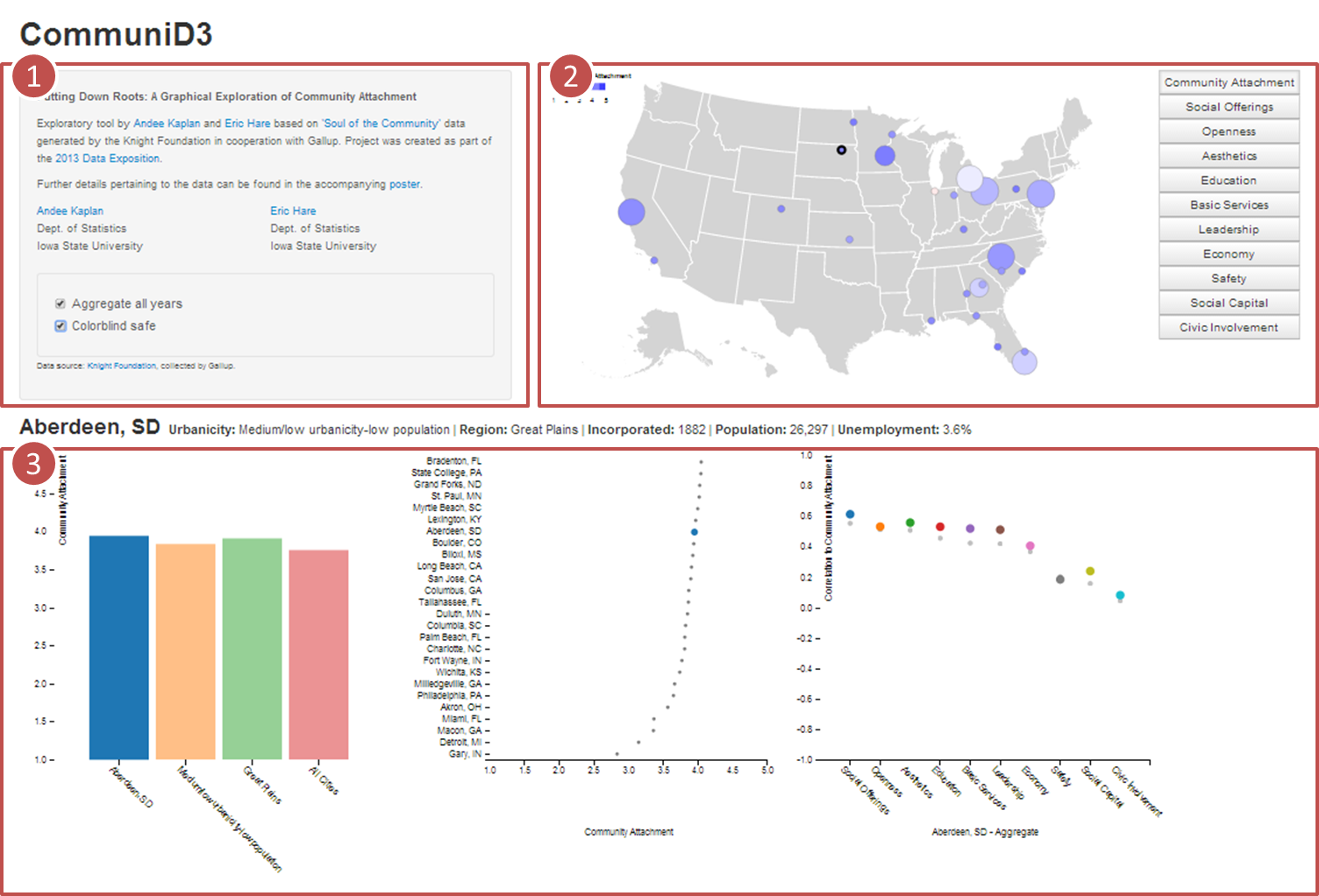}
\caption{\label{fig:tool} The components that make up CommuniD3, (1) Side panel, (2) Map panel, and (3) Plot panel.}
\end{figure}

CommuniD3 is comprised of three parts, \begin{inparaenum}[(1)]
\item Side panel, 
\item Map Panel, and
\item Plot panel,
\end{inparaenum}
as seen in Figure~\ref{fig:tool}.
As the user interacts with each piece, the other portions of the interface update to reflect the interaction. In this way we have built an interactive graphical framework, rather than an animation.

\begin{figure}[H]
\centering
\includegraphics[width=\textwidth]{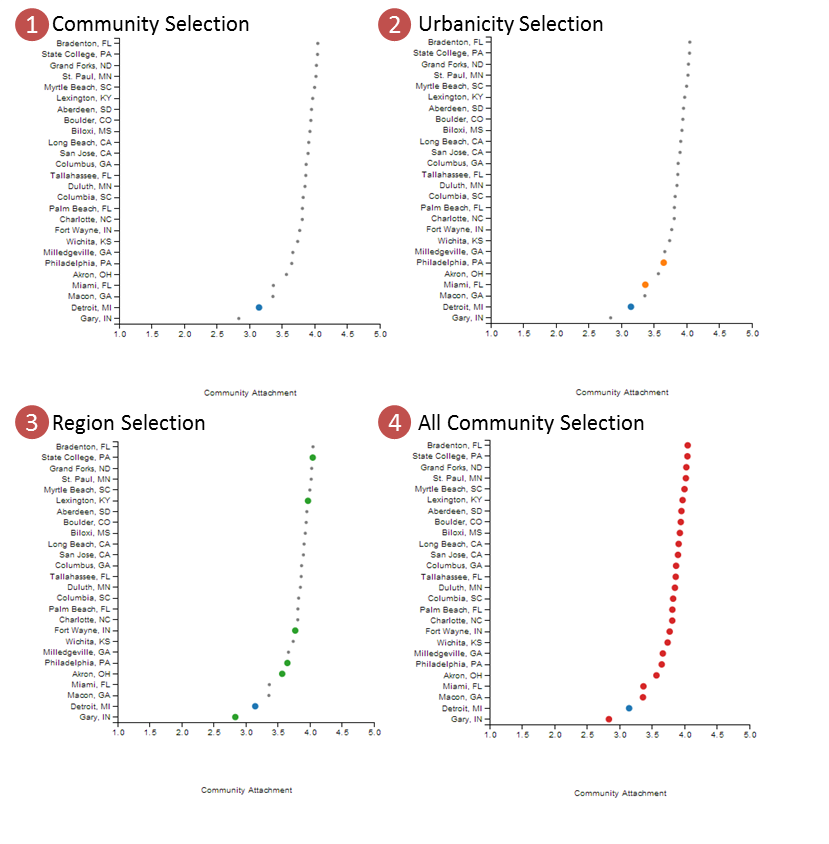}
\caption{\label{fig:dots} Examples of the different types of highlighting available in the ordered dot plot from the plot panel. The highlighting corresponds to (1) community selection, (2) urbanicity selection, (3) region selection, and (4) all community selection. In this example, Detroit, MI has been selected to display the values of community attachment for all three years, 2008-2010.}
\end{figure}

\begin{enumerate}[(1)]
\item
The {\bf side panel} houses two features. The first is the ability to investigate the data for individual years versus aggregated across all three years. In this way we are able to explore attitude changes across the three years surveyed as well as overarching trends in the regions and urbanicities. The second is a colorblind friendly option that uses a red-blue color scheme on the map rather than red-green to accommodate more users.
\item
The {\bf map panel} is the central piece of the application. The 26 communities surveyed are plotted geographically on a map of the United States. The size of each dot represents the number of surveys for the chosen time period, while the color shading of each dot corresponds to the average value for each community in the chosen time period for the metric selected. The panel on the right allows the user to change the metric displayed. Additionally, each community is clickable. On click, basic information about that community is displayed below the map panel and the plot panel is updated to reflect the community that is clicked. It is our goal for a community planner to be able to start with the map panel and find a community that was surveyed that corresponds to the community they are interested about, or one that is nearby, as a means to delve into the driving factors of community attachment.
\item
The {\bf plot panel} is a set of three linked plots that detail three aspects of the dataset. The first plot is a bar graph showing the average value of the metric selected for the year range selected for the community selected as well as for its urbanicity and its region. For example, once Detroit, MI is the community in focus, the urbanicity is ``Very high urbanicity-very large population'' and the region is the Rust Belt. The fourth bar in the chart represents the aggregation of all the communities serves as a reference. While the bar chart is a plot that shows surface information, its true purpose here is to control the information displayed in the other two plots. As the user clicks on the bars, the other two plots display information pertaining to the level selected (either region, urbanicity, or the whole dataset). The middle plot is an ordered dot plot displaying the average value for the metric selected for all communities with the level selected in the bar chart highlighted. See Figure~\ref{fig:dots} for an example of the different levels of highlighting. The third plot is a plot of pairwise correlations between each metric and community attachment for the level of aggregation (year and community/region/urbanicity). The small grey dots serve as a reference level in the background that displays the correlation for every survey aggregated to ease comparison for the user. The three plots are linked in such a way that selection through clicking in one plot will affect all three plots and potentially the map panel. In this manner, the user can truly drive their experience and take ownership of their analysis.
\end{enumerate}

\subsection{The Shoulders of Giants}
\begin{figure}[H]
\centering
\includegraphics[width=\textwidth]{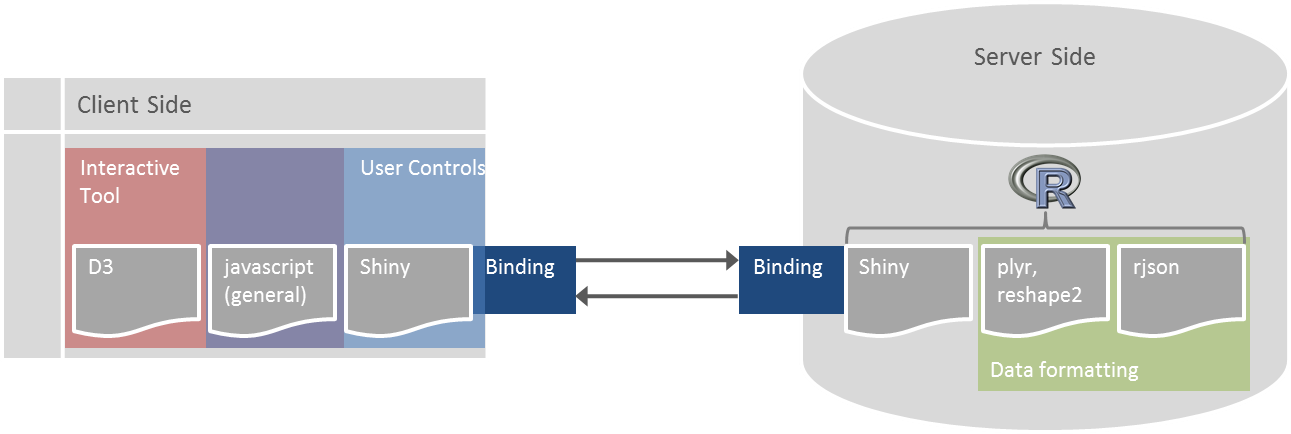}
\caption{\label{fig:D3shiny} Diagram of the uses of \texttt{D3} and \texttt{Shiny} in CommuniD3, specifically focusing on client versus server utilization. \texttt{D3} and JavaScript are used to create the visualizations, as well as control the user interaction. \texttt{Shiny} and the associated {\tt R} packages are the framework on which the application is built.}
\end{figure}

We incorporated several pioneering technologies in the creation of our application that allowed us to find insights in the dataset. \texttt{Shiny} \citep{rs-shiny} is an {\tt R} package created by RStudio that enables {\tt R} users to create an interactive web application that utilizes {\tt R} as the background engine. In CommuniD3, \texttt{Shiny} is used as the framework upon which the application sits. \texttt{D3} \citep{mb-d3} stands for ``Data Driven Documents" and is a JavaScript library developed and maintained by Mike Bostock with the purpose of visualizing and interacting with data in a web-based interface. We used \texttt{D3} and JavaScript to create the visualizations as well as to control all the user interaction with the application. The graphics and user interface are all stored entirely on the client side, allowing for seamless transitions of the graphics. See Figure~\ref{fig:D3shiny} for a diagram of the ways \texttt{Shiny} and \texttt{D3} are used in CommuniD3.

We also leveraged other {\tt R} packages to help with data manipulation. We used \texttt{plyr} \citep{plyr}, \texttt{reshape2} \citep{reshape2}, and \texttt{rjson} \citep{rjson} to split and aggregate metric values according to the levels selectable by the user before passing the data to the client side in the JSON format. For subsequent analysis after using CommuniD3 we used the {\tt R} packages \texttt{ggplot2} \citep{ggplot2} and \texttt{maps} \citep{maps} to dive deeper into the interesting findings from the application.

\section{Stories}

In this section, we illustrate a specific example of how CommuniD3 can be used to highlight interesting features in the data. We then proceed by showing other interesting findings in the data, aided by the use of CommuniD3. 

We elected to divide our analysis using two primary factors. The first was the geographic region the community was located in, and the second was the urbanicity of the particular community. Urbanicity is a census designation which was provided in the dataset, while regions were determined by us. The interactive tool was then used to help us discover a story in the data for each of the five regions.

\subsection{Great Plains - Effect of Quality Education}

\begin{figure}[H]
\centering
\includegraphics[width=\textwidth]{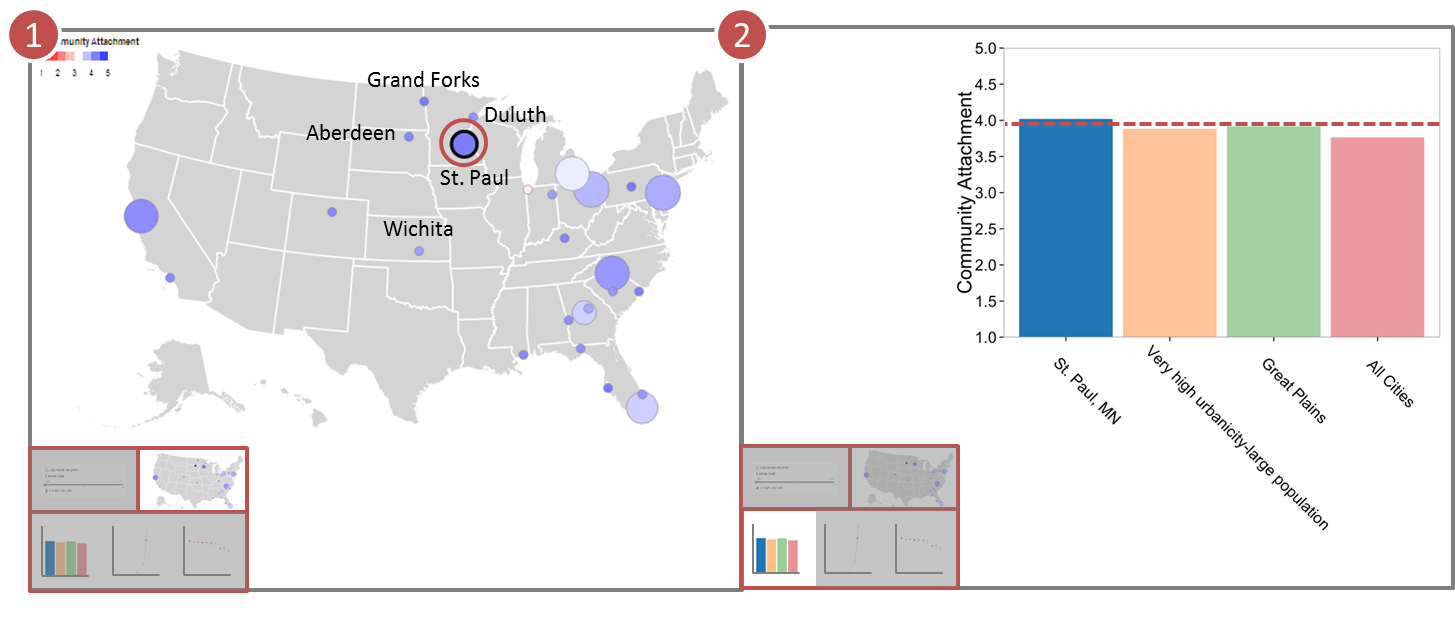}
\caption{\label{fig:greatplains_1} (1) Select Saint Paul  (2) Focus on the comparison of Community Attachment in Saint Paul with other cities that have geographical or cultural similarities. Conclude that Saint Paul is more attached than most of its counterparts.}
\end{figure}

\begin{figure}[H]
\centering
\includegraphics[width=\textwidth]{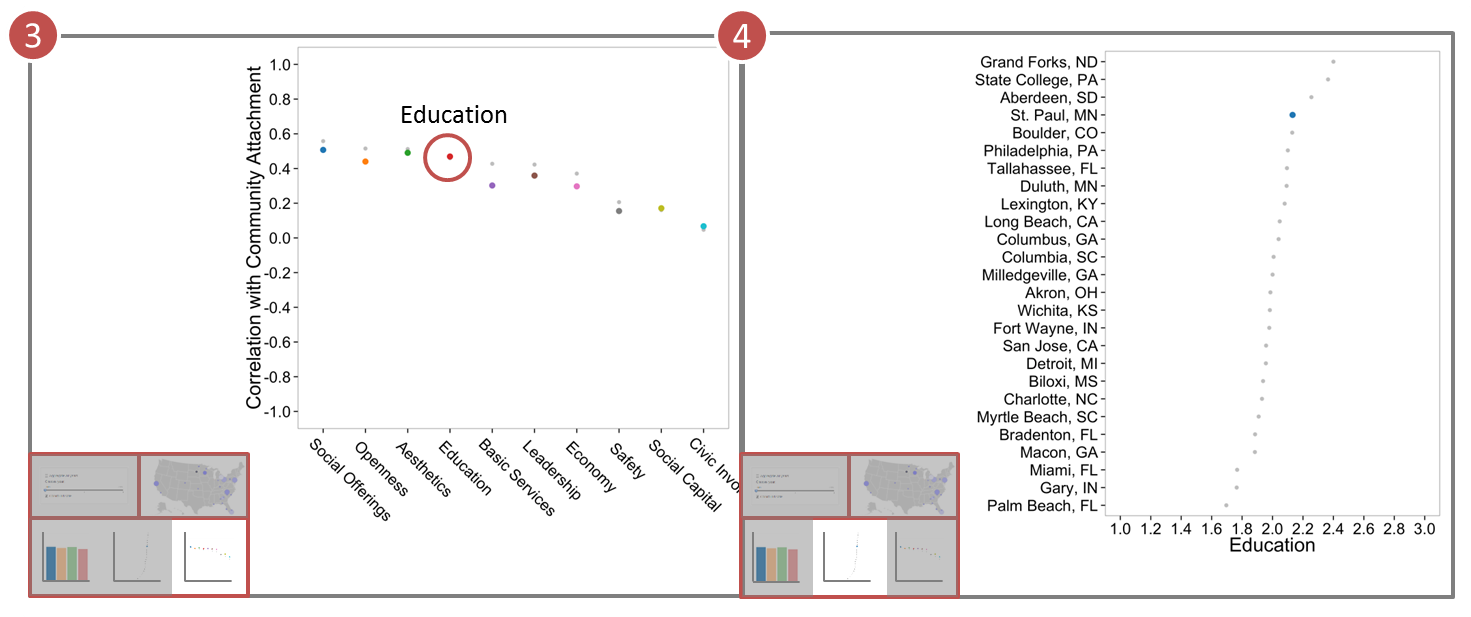}
\caption{\label{fig:greatplains_2} (3) Focus on the correlation between Community Attachment and the other metrics in Saint Paul to identify Education as particularly important.  (4) Compare the mean value for education in Saint Paul with the other cities.}
\end{figure}

\begin{figure}[H]
\centering
\includegraphics[width=\textwidth]{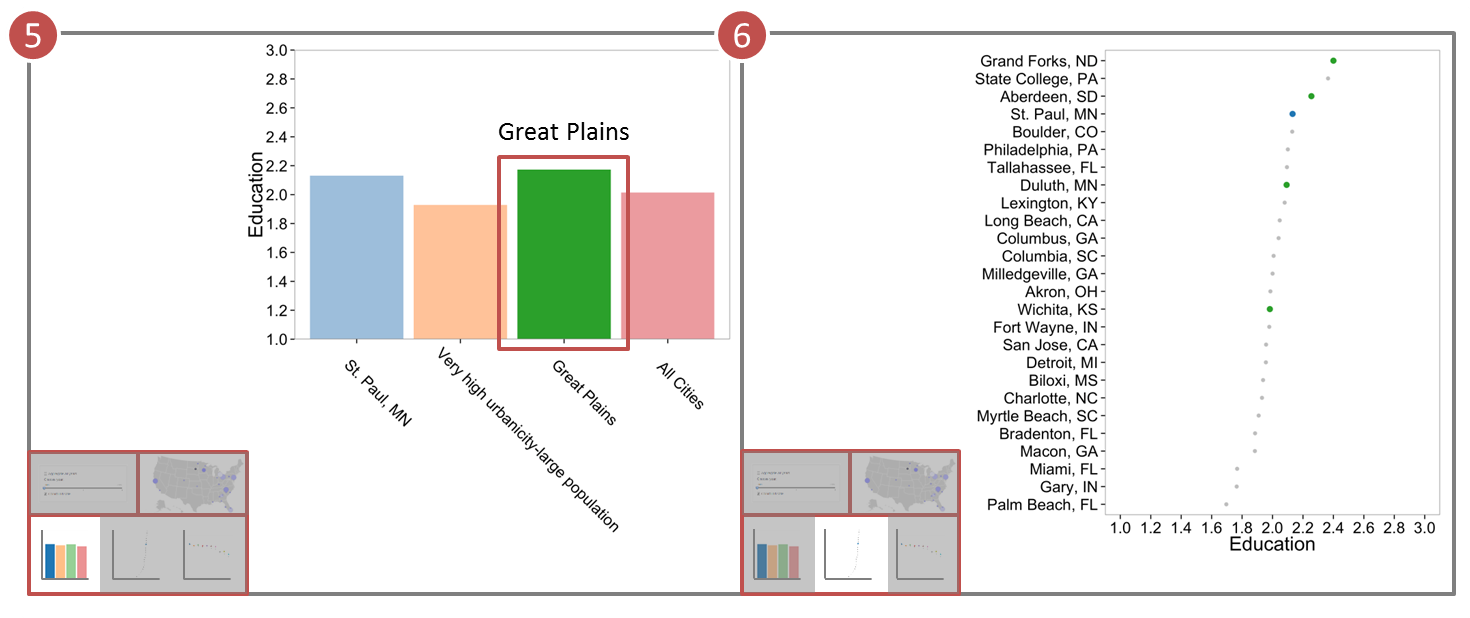}
\caption{\label{fig:greatplains_3} (5) Focus on how the mean value for Education is larger than the mean value for all communities aggregated.  (6) Examine the comparison of all cities to observe that three of the top four communities in Education are located in the Great Plains.}
\end{figure}

The five communities comprising the Great Plains were Grand Forks, ND, Duluth, MN, Aberdeen, SD, Saint Paul, MN, and Wichita, KS. Through use of CommuniD3, we quickly discovered that the individuals in this region rated the quality of education in the community more highly. Figures~\ref{fig:greatplains_1}, \ref{fig:greatplains_2}, and \ref{fig:greatplains_3} illustrate the sequence of steps in CommuniD3 which led to that conclusion.

You recommend wish to follow along in CommuniD3 as we describe the steps taken. First, we click on Saint Paul, as shown in panel (1). We examine the bar charts of Community Attachment displayed in panel (2), which indicates the Great Plains region has an overall higher attachment than the average of all other cities.

We can then look at the plot of correlations of panel (3), demonstrating that the Great Plains communities have an overall larger correlation between Education and Community Attachment than the average of all communities. This leads to switching the metric of interest in CommuniD3 to Education, and an examination of the ordered dot plot of panel (4). This plot makes clear that while Saint Paul is strong in education (ranking fourth of all communities), three other communities particularly stand out.

At this point, the story of the Education metric in the Great Plains becomes clear. Panel (5) illustrates the bar chart that was first shown in panel (2), but this time with regards to Education. Saint Paul, despite ranking fourth overall, actually has a slightly lower value for this metric than does the Great Plains communities aggregated. This is explained by the fact that the Great Plains communities stand above the rest, averaging 2.2 out of 3 in Education compared to about 2.0 for the average of all cities. Our final step is to click on the Great Plains bar, which immediately highlights the Great Plains communities in the ordered dot plot as shown in panel (6).

It is quickly evident that Grand Forks, Aberdeen, and Saint Paul comprise three of the top four communities in the Education metric, helping to explain the fact that the Great Plains has the overall largest Community Attachment among the regions we considered.

A similar approach was taken for the other four regions, leading to the conclusions presented in this paper. Henceforth, we illustrate the findings in static graphics and tables for ease of presentation, all of which were motivated and discovered through use of CommuniD3.

\subsection{West - Urbanicity and Openness}
Another focus of our analysis was the influence of the Urbanicity designations of the communities. In the West region, which comprises Boulder, CO, San Jose, CA, and Long Beach, CA, we found some evidence of an urbanicity-specific metric correlated with attachment.

Table \ref{tbl:open_table} displays the top five communities by the Openness Metric. First, it can be seen that the three communities comprising the West Region were all in the top five for Openness. Second, Boulder and Long Beach each have the urbanicity of "Very high urbanicity-medium population". These communities consist of a relatively modest population, but where most of whom live in the urban core of the city. Figure \ref{fig:west_one} displays a two-dimensional bin plot of Community Attachment versus Openness, displaying only the communities with the designation "Very high urbanicity-medium population". Areas of darker red have a higher frequency of responses than those that are white. Notice that both Boulder and Long Beach had more respondents indicating a high Community Attachment and a high Openness. By comparison, Akron, OH, and Gary, IN, two communities in the Rust Belt region with the same urbanicity, had much lower ratings on both of these scales. Bradenton, FL had many citizens highly attached to the community, but somewhat lower ratings for Openness compared to Boulder and Long Beach. Ultimately, communities in the West region with this urbanicity designation placed a higher value on the Openness of their community than do other communities of similar size in the rest of the country.

\begin{table}[ht]
\centering
\begin{tabular}{lllr}
  \hline
Community & Region & Urbanicity & Openness \\ 
  \hline
Long Beach, CA & West & Very high urbanicity-medium population & 1.95 \\ 
  San Jose, CA & West & Very high urbanicity-large population & 1.88 \\ 
  St. Paul, MN & Great Plains & Very high urbanicity-large population & 1.88 \\ 
  State College, PA & Rust Belt & Medium/low urbanicity-low population & 1.87 \\ 
  Boulder, CO & West & Very high urbanicity-medium population & 1.84 \\ 
   \hline
\end{tabular}
\caption{Top Five Communities by the Openness Metric. Note that three of the communities are from the West region} 
\label{tbl:open_table}
\end{table}

\begin{knitrout}
\definecolor{shadecolor}{rgb}{0.969, 0.969, 0.969}\color{fgcolor}\begin{figure}[H]

{\centering \includegraphics[width=\maxwidth]{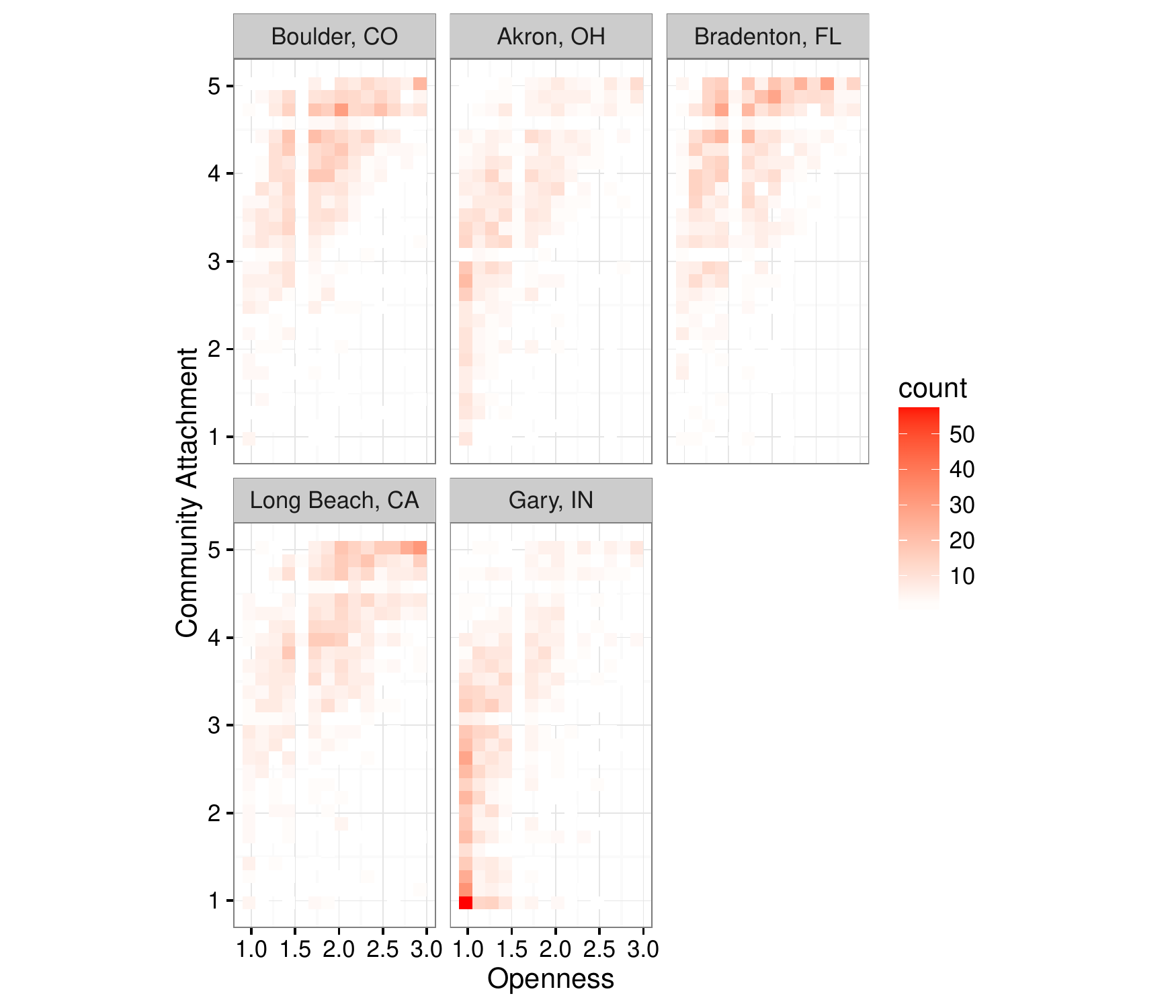} 

}

\caption[2D binned plot of responses for Openness and Community Attachment among the five communities with Very high urbanicity-medium population designations]{2D binned plot of responses for Openness and Community Attachment among the five communities with Very high urbanicity-medium population designations. Communities in the West region with this urbanicity designation placed a higher value on the Openness of their community than do other communities of similar size in the rest of the country.}\label{fig:west_one}
\end{figure}

\end{knitrout}

\subsection{Deep South - The Safety Angle}
Exploring trends in the Deep South communities of Macon, GA, Milledgeville, GA, Columbus, GA, Tallahassee, FL, and Biloxi, MS quickly suggested that residents of these communities were displeased with the Safety of the community. In 2008, Macon and Columbus ranked third and fourth worst respectively among all 26 communities in terms of Safety. By 2010, the situation degraded further, as Macon declined to the overall worst Safety rating, while Milledgeville ranked third worst, and Columbus remained the fourth worst. Biloxi was a notable exception, however, ranking eighth best in 2010. Biloxi also exceeded its fellow Deep South communities in terms of Social Offerings, ranking second best amongst all communities in each of the three years the survey was conducted, and by far the best amongst the communities in the Deep South.

As it turns out, Biloxi also had the highest overall Community Attachment rating in the Deep South. Figure \ref{fig:ds_one} displays the average rating from 2008 to 2010 in terms of Safety, Social Offerings, and Community Attachment. Biloxi is highlighted in red, the other Deep South communities are highlighted in blue, and the rest of the communities are highlighted in gray. The stark difference between Biloxi and the rest of the Deep South is readily apparent, and helps to explain why Community Attachment was quite high in Biloxi.

\begin{knitrout}
\definecolor{shadecolor}{rgb}{0.969, 0.969, 0.969}\color{fgcolor}\begin{figure}[H]

{\centering \includegraphics[width=\maxwidth]{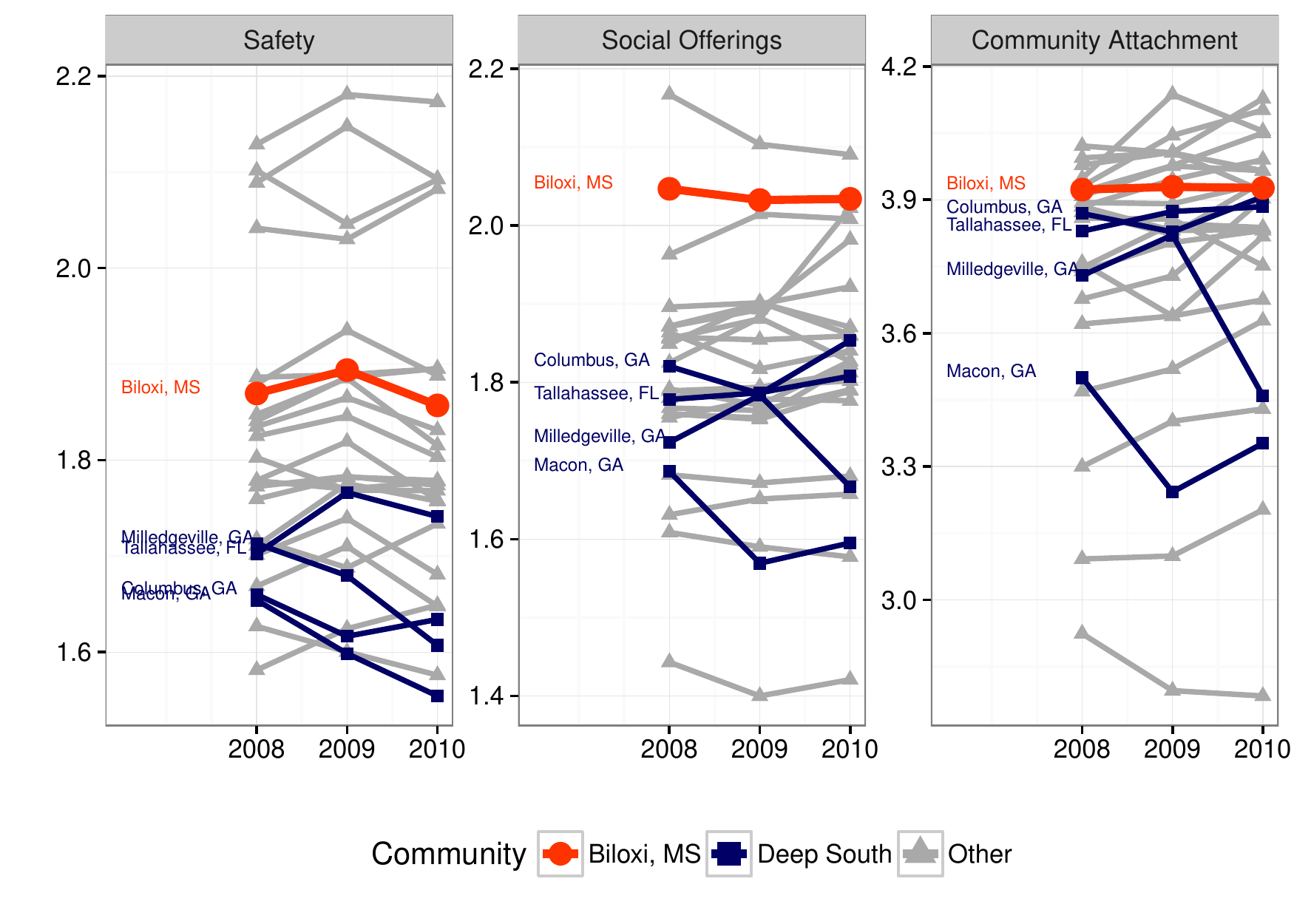} 

}

\caption[Responses across the three years for Safety, Social Offerings, and Community Attachment]{Responses across the three years for Safety, Social Offerings, and Community Attachment. The stark difference between Biloxi and the rest of the Deep South in terms of Safety and Social Offerings is readily apparent. This helps to explain why Community Attachment was quite high in Biloxi.}\label{fig:ds_one}
\end{figure}

\end{knitrout}

\subsection{Southeast - Social Offerings in Myrtle Beach}
In the Deep South, we saw some evidence suggesting Biloxi's high rating for Social Offerings may have contributed to a strong sense of attachment in that community. Nowhere is this phenomenon more prominent than in the Southeast region, where Myrtle Beach, SC is located. Myrtle Beach was the fifth most attached community amongst all communities in the dataset. However, Myrtle Beach did no better than 13th in all other metrics with the exception of Social Offerings, where it was ranked first. In other words, residents of Myrtle Beach felt the Social Offerings in their community were very strong, while most other metrics, including Aesthetics, Openness, Safety, and Education were poor. Figure \ref{fig:southeast_one} illustrates this phenomenon with a parallel coordinate plot. The mean value for all metrics is displayed for each of the communities, with Myrtle Beach highlighted in green, and other Southeast communities highlighted in blue. The metrics are sorted from high to low by the average value for each metric amongst all the communities. Notice that Myrtle Beach fairly closely tracked the rest of the communities in all metrics except for Social Offerings, where a sizable ``jump'' can be observed.

\begin{knitrout}
\definecolor{shadecolor}{rgb}{0.969, 0.969, 0.969}\color{fgcolor}\begin{figure}[H]

{\centering \includegraphics[width=\maxwidth]{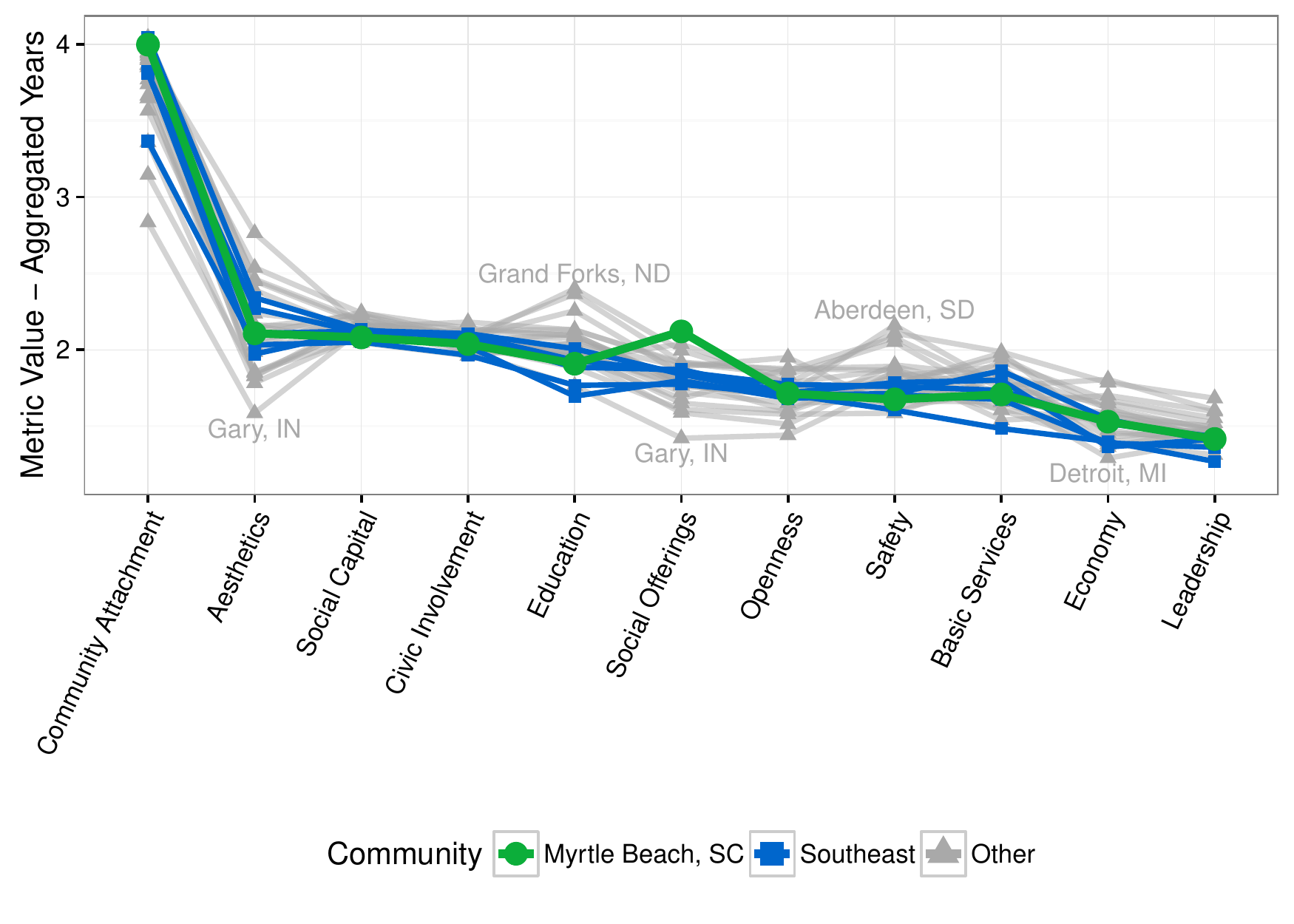} 

}

\caption[Mean value of metric for each community]{Mean value of metric for each community. Myrtle Beach is highlighted in green, and the other communities in the Southeast are highlighted in blue. Myrtle Beach stands closely tracked the rest of the communities in all metrics except for Social Offerings, where they were the highest rated.}\label{fig:southeast_one}
\end{figure}

\end{knitrout}

The question remains of how Myrtle Beach had such highly attached citizens when only Social Offerings appeared to be a positive metric for the community. The reason is that Social Offerings was the single most highly correlated metric with Community Attachment in 23 out of the 26 communities, including Myrtle Beach.

\subsection{Rust Belt - The Economic Collapse}
As the data covered the period from 2008 to 2010, we hoped to find some stories relating to the economic collapse, or the Great Recession, which began in 2008. We focused on the Economy metric in the Rust Belt communities, and noted a largely negative view of economic conditions in this region, particularly in its economic center, Detroit, MI. Table \ref{tbl:econ} displays the average response on a 1-3 scale for Economy in the Rust Belt communities across each of the three years. It can be seen that although all Rust Belt communities experienced a drop in attitude about the economy between 2008 and 2009, Detroit's was noticeably smaller. By 2010, attitudes about the economy began to improve. This can also be seen by examining Figure \ref{fig:rb_one}, a density plot of values for Economy in each of the communities in the three years.

\begin{table}[ht]
\centering
\begin{tabular}{lrrr}
  \hline
Community & 2008 & 2009 & 2010 \\ 
  \hline
Detroit, MI & 1.26 & 1.25 & 1.37 \\ 
  Gary, IN & 1.50 & 1.28 & 1.37 \\ 
  Akron, OH & 1.41 & 1.32 & 1.41 \\ 
  Fort Wayne, IN & 1.50 & 1.36 & 1.51 \\ 
  Philadelphia, PA & 1.60 & 1.42 & 1.49 \\ 
  Lexington, KY & 1.69 & 1.53 & 1.64 \\ 
  State College, PA & 1.65 & 1.59 & 1.72 \\ 
   \hline
\end{tabular}
\caption{The average value for the Economy metric in the five Rust Belt communities in each of the three years, sorted by the aggregated mean value for all three years. Notice the large drop in economic outlook for all cities in the Rust Belt from 2008 to 2009. Detroit, however, dropped only 0.01 in average economic outlook in that timespan.} 
\label{tbl:econ}
\end{table}

\begin{knitrout}
\definecolor{shadecolor}{rgb}{0.969, 0.969, 0.969}\color{fgcolor}\begin{figure}[H]

{\centering \includegraphics[width=\maxwidth]{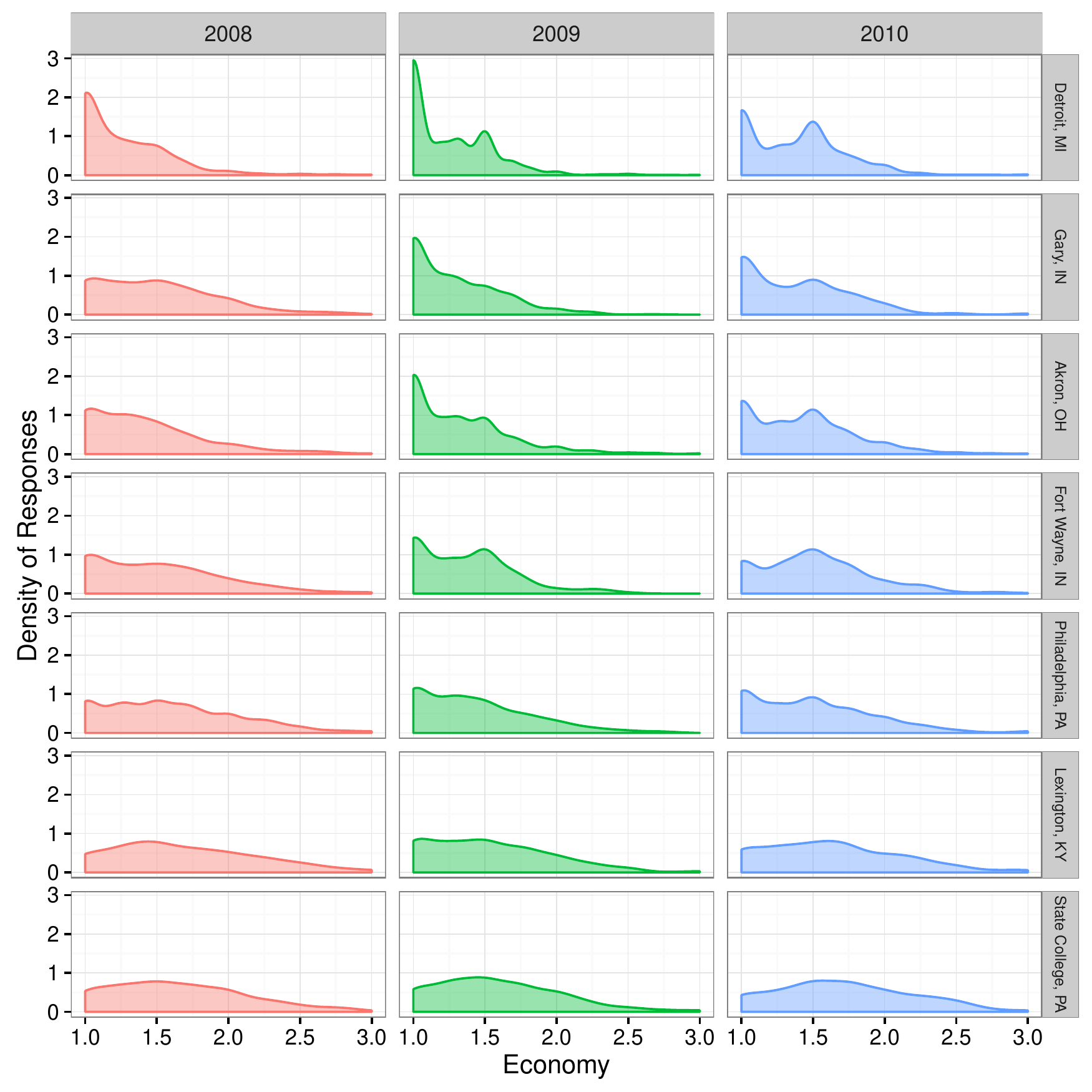} 

}

\caption[Density of responses for Economy in Rust Belt communities over each of the three years]{Density of responses for Economy in Rust Belt communities over each of the three years. It appears that Detroit experiences a modest recovery in terms of economic outlook from 2009 to 2010 that the other communities in the Rust Belt did not experience.}\label{fig:rb_one}
\end{figure}

\end{knitrout}

What this table and plot suggests is that although it is widely believed that Detroit was hit especially hard by the economic collapse, there was a bit of resillience in the 2009 and 2010 time frame. Attitudes in Detroit were low regarding the economy in 2008, but did not exhibit much worsening in 2009, and began to improve in 2010. Perhaps the automotive bailouts, which were passed and signed into law in 2009, may be a reasonable explanation for why this occurred in America's ``Motor City".

\section{Conclusion}

CommuniD3, with its use of linked plots and user interaction, allows for the discovery of features and trends in the data that are far more difficult to discover with static plots alone. By creating the tool prior to analyzing the data, we were able to find stories in the data that may not have been as readily apparent otherwise. Specifically, we uncovered an especially strong link between quality of education and Community Attachment in the Great Plains, the importance of Social Offerings in Myrtle Beach and Biloxi, and slowly recovering economic attitudes in the Rust Belt.

While the interactive tool is specialized for this particular dataset, the philosophy and ideas behind its creation hold for other data and applications. By empowering the user to guide his or her own discoveries, the analysis of data can be completed by subject-matter experts in their fields who may be less technologically inclined. The flexibility and ease of \texttt{Shiny}, combined with the interactivity of \texttt{D3} will hopefully open up a whole new set of possibilities for analyzing complex datasets more easily.

\clearpage

\printbibliography
\end{document}